\documentclass[aps, prd, reprint, longbibliography, nofootinbib,superscriptaddress, floatfix]{revtex4-2}
\pdfoutput=1

\usepackage{amsmath,amssymb,amsfonts}
\usepackage{mathrsfs}
\usepackage{bbm}
\usepackage{slashed}
\usepackage{graphicx}
\usepackage{comment}
\usepackage{verbatim}
\usepackage[T1]{fontenc} 
\usepackage[utf8]{inputenc}
\usepackage[colorlinks]{hyperref}
\usepackage{mathbbol}
\usepackage[dvipsnames]{xcolor}
\usepackage{svg}
\usepackage{flushend}
\usepackage{physics}
\usepackage{multirow}
\usepackage{graphicx}
\usepackage{dcolumn}
\usepackage{bm}
\usepackage{tensor}
\usepackage{comment}
\usepackage[utf8]{inputenc} 
\usepackage{graphicx,color,overpic,mathtools}
\usepackage{amsthm,amsmath,amssymb,hyperref,mathrsfs}
\usepackage{braket,bm,bbm}
\usepackage{cancel}
\usepackage{physics}
\usepackage{float}
\usepackage[english]{babel}
\usepackage{graphicx}
\hypersetup{
    colorlinks=false,
    pdfborder={0 0 0},
}

\usepackage[normalem]{ulem}

\newcommand{\qu}[1]{``{#1}''} 

\begin{document}

\title{Hidden wormholes from an action principle in $D=4$}

\author{Francesco Del Porro}
\email[]{francesco.del.porro@nbi.ku.dk}
\affiliation{Center of Gravity, Niels Bohr Institute, Blegdamsvej 17, DK-2100 Copenhagen \O, Denmark}
\affiliation{Niels Bohr International Academy, Niels Bohr Institute, Blegdamsvej 17, DK-2100 Copenhagen \O, Denmark}

\author{Alessia Platania}
\email[]{alessia.platania@nbi.ku.dk}
\affiliation{Center of Gravity, Niels Bohr Institute, Blegdamsvej 17, DK-2100 Copenhagen \O, Denmark}
\affiliation{Niels Bohr International Academy, Niels Bohr Institute, Blegdamsvej 17, DK-2100 Copenhagen \O, Denmark}
\affiliation{Institute of Physics, University of Graz, Universit\"atsplatz 5, 8010 Graz, Austria}

\author{Samuele Silveravalle}
\email[]{ssilvera@sissa.it}
\affiliation{SISSA - International School for Advanced Studies, Via Bonomea 265, 34136 Trieste, Italy}
\affiliation{INFN, Sezione di Trieste, Via Valerio 2, 34127 Trieste, Italy}
\affiliation{IFPU - Institute for Fundamental Physics of the Universe, Via Beirut 2, 34151 Trieste, Italy}

\begin{abstract}
We present new black hole solutions for the class of gravitational effective field theories in four spacetime dimensions that extend General Relativity while retaining the graviton as the only degree of freedom of gravitational origin. We consider terms up to cubic order in curvature, solve the resulting field equations, and find one-way hidden wormholes as numerical solutions: these are regular black holes with an interior wormhole throat that is crossable in one direction and leads to a cosmological spacetime in its interior. They are regular and have no inner horizon, thus evading the mass inflation instability that afflicts conventional regular black hole geometries.
\end{abstract}

\maketitle

\paragraph*{Introduction.}

Black holes provide a remarkable arena for testing the strong-field regime of gravity and exploring the implications of its possible ultraviolet (UV) completion. In General Relativity (GR), gravitational collapse generically leads to spacetime singularities, indicating a breakdown of the classical theory~\cite{Penrose:1964wq,Hawking:1970zqf}. 

Several regular alternatives have been constructed that avoid singular behaviors of classical solutions in GR (see~\cite{Buoninfante:2024oxl,Carballo-Rubio:2025fnc} for reviews). They violate the energy conditions and replace the inner singularity with a regular core. Such geometries can exhibit a variety of interior structures, including de Sitter cores and wormhole throats connecting distinct spacetime regions. Despite their appealing regularity properties, many such geometries face non-trivial consistency and stability challenges.
For the most common regular black holes~\cite{Dymnikova:1992ux,Hayward:2005gi,Bambi:2013ufa}, for instance, regularity comes at the price of introducing an inner horizon, which is the doorway to mass inflation instabilities~\cite{Poisson:1989zz,Ori:1991zz,Carballo-Rubio:2018pmi,Carballo-Rubio:2026gwg}. Extremal configurations are subject to the Aretakis instability~\cite{Carballo-Rubio:2022nuj}, tidal force divergences~\cite{Horowitz:2023xyl,Horowitz:2024dch,Cano:2025ejw,DelPorro:2025fiu}, and lead to problematic remnants~\cite{Susskind:1995da}. Rapidly rotating horizonless objects can develop ergoregion instabilities~\cite{Cardoso:2007az,Maggio:2018ivz,Franzin:2022iai,Maggio:2017ivp}. Stable light rings, which are generically present in sufficiently compact objects, may trigger nonlinear instabilities~\cite{Cardoso:2014sna,Cunha:2017qtt}. Physical surfaces and wormhole throats replacing outer horizons can also be unstable against accretion~\cite{Carballo-Rubio:2025fnc}. At the same time, the semiclassical collapse dynamics could affect the formation of singularities~\cite{Gralla:2025gzl,DiFilippo:2025kzh,DelPorro:2026njz}. In addition to such stability issues, the physical relevance of regular black-hole-like geometries depends not only on their kinematic properties but also on their dynamical origin from a principle of least action. In particular, a metric constructed \textit{ad hoc} to satisfy regularity conditions need not arise from an action principle~\cite{Knorr:2022kqp}.

In this context, there is a more solid way to ask what lies beyond GR. If the graviton is the only degree of freedom of gravitational origin, the leading deviations from GR are encoded in an effective field theory (EFT) of the metric~\cite{Donoghue:1994dn,Burgess:2003jk}: the most general diffeomorphism-invariant action of one propagating massless graviton. Any consistent quantum gravity theory is expected to reduce to such an EFT at low energies, with the Wilson coefficients set by the specific UV completion~\cite{Basile:2024oms,Buoninfante:2024yth}. The spectrum of solutions of this EFT is thus a prediction common to all such completions, and is paramount to determining which black-hole-like solutions replace Schwarzschild black holes in the full theory.
In such an EFT framework, the leading correction to GR that cannot be absorbed in a field redefinition is the Goroff-Sagnotti operator~\cite{Goroff:1985sz}, i.e., the curvature invariant built out of the full contraction of three Weyl tensors. Importantly, invariants constructed from the Weyl tensor encode interactions that are not captured by the Ricci-flat Einstein equations alone. Although the implications of higher-curvature corrections to black hole physics have been extensively studied in higher dimensions~\cite{Myers:1986un,Kanti:2004nr,Emparan:2008eg,Emparan:2014wki,Emparan:2014aba,Andrade:2018nsz,Andrade:2018rcx,Andrade:2019edf,Emparan:2025wsh}, in particular in quasi-topological theories~\cite{Bueno:2024dgm,Bueno:2024dgm,Bueno:2024zsx,Bueno:2024eig,Bueno:2026dln,Bueno:2026oyg,DiFilippo:2026puc,Cano:2026uch}, obtaining exact or controlled interior geometries directly from the effective field equations in $D=4$ remains a nontrivial task. First attempts in this direction have been taken in~\cite{Anselmi:2013wha,deRham:2020ejn,Knorr:2022kqp,Daas:2023axu}, while solutions in more general quadratic and cubic truncations have been classified in~\cite{Pravda:2020zno,Bonanno:2019rsq,Silveravalle:2022wij,Giacchini:2025gzw,Giacchini:2025mlv,Borissova:2025nvj}.

In this Letter, we consider the minimal EFT beyond GR, namely the Einstein-Hilbert action supplemented by the aforementioned Goroff-Sagnotti operator, and we look for its static, spherically symmetric solutions. We solve the modified field equations and obtain regular black hole geometries whose interiors contain a wormhole throat connecting to a cosmological spacetime. The throat is traversable in only one direction, and the solutions possess no inner horizon, thereby avoiding the standard mass inflation mechanism associated with such horizons, as well as the typical instabilities of horizonless objects~\cite{Lobo:2020ffi}. Our results establish that these geometries can arise directly from higher-curvature gravitational dynamics, providing an explicit realization of regular black hole interiors within a systematic EFT description of gravity in four spacetime dimensions.

\paragraph*{Gravitational EFTs up to cubic order.}

A systematic and model-independent way to parametrize corrections beyond GR is the framework of EFT~\cite{Donoghue:1994dn,Burgess:2003jk}. In pure gravity, operators related by field redefinitions are physically redundant, so the EFT is naturally expressed in terms of essential operators. In $D=4$, the leading correction beyond GR arises at cubic order in the curvature and is given by the Goroff-Sagnotti operator~\cite{Goroff:1985sz},
\begin{equation}
    C^3\equiv C_{\mu\nu}{}^{\rho\sigma}C_{\rho\sigma}{}^{\alpha\beta}
C_{\alpha\beta}{}^{\mu\nu}\,.
\end{equation}
A gravitational EFT is thus expanded as~\cite{Metsaev:1986yb,Donoghue:1994dn,Burgess:2003jk}
\begin{equation} \label{eq:action}
    S=\frac{1}{16\pi G}\int {\rm d}^4x\sqrt{-g}\,\Big[R+\frac{\lambda}{8 m_{\text{Pl}}^4}\, C^3+\dots \Big]\,,
\end{equation}
with $m_{\text{Pl}}=G^{-1/2}$. Here we set the cosmological constant to zero, which is a good approximation at the scales of interest. Higher-order operators are suppressed by increasing powers of the Planck scale. The form of this effective action is universal: it is fixed entirely by the field content and symmetries, and is shared by all theories that reduce to GR at low energies,  without additional degrees of freedom. The UV completion enters only through the values of the Wilson coefficients. Thus, EFT determines which corrections to GR are possible, while the UV completion determines their size. In this context, it is key to determine how EFT-motivated corrections modify the spectrum of black-hole-like solutions and whether genuinely new solutions can arise. To this end, we shall focus on the impact of the leading-order correction to GR: the cubic operator and the associated Goroff-Sagnotti coupling.

\paragraph*{Equations of motion and numerical solutions.} \label{sec:Numerics}

The full equations of motion (EOM) are lengthy and not particularly informative. They follow from varying the Einstein-Hilbert and Goroff-Sagnotti terms, and their explicit form can be found in the literature, \textit{e.g.} in~\cite{Alvarez:2023gfg}.

We investigate static and spherically symmetric solutions, starting with the most general ansatz for a metric in Schwarzschild coordinates
\begin{equation} \label{eq:static_patch}
   {\rm d}s^2=- h(r) {\rm d}t^2 + \frac{{\rm d}r^2}{f(r)} + r^2 {\rm d} \Omega_2^2\,,
\end{equation}
where  ${\rm d}\Omega_2^2={\rm d} \theta^2+\sin^2(\theta){\rm d}\phi^2 $ is the volume element on the 2-sphere. This parametrization allows us to identify relevant regions of spacetime: the asymptotically flat region corresponds to $h(r),f(r)\to1$ as $r\to\infty$, an event horizon located at $r_{\scriptstyle H}$ is defined by $h(r_{\scriptstyle H})=f(r_{\scriptstyle H})=0$, while a wormhole throat at $r_{\scriptstyle T}$ satisfies $f(r_{\scriptstyle T})=0$ but $h(r_{\scriptstyle T})\neq0$~\cite{Morris:1988cz}. In particular, in the asymptotically flat region the metric is completely characterized by the ADM mass $M$ and the Goroff-Sagnotti coupling $\lambda$~\cite{Anselmi:2013wha,deRham:2020ejn,Knorr:2022kqp,Daas:2023axu}.

For the numerical construction of solutions, we use conformal-to-Kundt coordinates~\cite{Pravda:2016fue,Svarc:2018coe,Podolsky:2018pfe,Podolsky:2019gro,Pravda:2020zno,Daas:2023axu}, in which the metric takes the form
\begin{equation} \label{eq:conf_kundt}
  {\rm d}s^2=\Omega^2(x)\left(\mathcal{H}(x) {\rm d}v^2 -2{\rm d}v{\rm d}x + {\rm d} \Omega_2^2 \right)\,.
\end{equation}
In the region covered by the original chart, the new variables are related to $(t,r)$ through
\begin{equation}\label{eq:ctk_variables}
v=t+\int\frac{{\rm d}r}{\sqrt{f(r)h(r)}}\,, \qquad
x=-\int\frac{{\rm d}r}{r^2}\sqrt{\frac{h(r)}{f(r)}}\,,
\end{equation}
where we used ingoing coordinates, and the metric functions are related to the original ones through $\Omega(x)=r(x)$ and $\mathcal{H}(x)=-r^{-2}(x)h(r(x))$. The integration constant in $x$ is chosen such that spatial infinity corresponds to $x\to0^+$.  
In this parametrization, asymptotic flatness corresponds to $\mathcal{H}(x)\sim-x^2$ and $\Omega(x)\sim x^{-1}$ as $x\to 0^+$, an event horizon is located at $x_{\scriptstyle H}$ when $\mathcal{H}(x_{\scriptstyle H})=0$ and $\Omega(x_{\scriptstyle H})\neq0$, while a wormhole throat is identified by a local minimum of $\Omega(x)$ at $x_{\scriptstyle T}$ with $\mathcal{H}(x_{\scriptstyle T})\neq0$ and $\Omega(x_{\scriptstyle T})\neq0$. Finally, the center of the geometry corresponds to $\Omega(x)=0$.

In conformal-to-Kundt coordinates, the EOM reduce to an autonomous system of two coupled ordinary differential equations~\cite{Daas:2023axu}. We construct their solutions using a shooting method between the asymptotically flat region and the event horizon. This procedure determines the free parameters of the solution near the horizon in terms of the ADM mass and the Goroff-Sagnotti coupling. The integration can then be extended across the event horizon to investigate the interior geometry. We use the coordinate $x$ throughout the numerical analysis and in the plots of the resulting solutions. Details of the EOM, the asymptotic and near-horizon expansions, and the numerical implementation are given in the Supplemental Material.

The numerical solutions are highly sensitive to the choice of parameters, particularly to the value of~$\lambda$ which enters both in the asymptotic expansion and in the EOM themselves. For this reason, we have not been able to systematically explore a large region of the parameter space, and we leave a more extensive investigation to future work. Nevertheless, for each value of~$\lambda$ considered, we find a threshold mass~$M_\lambda$ above which the solutions have an event horizon. More interestingly, whenever the numerical integration is sufficiently stable in the interior, we find a wormhole throat.

This specific feature is characteristic of geometries known as \textit{one-way hidden wormholes}~\cite{Carballo-Rubio:2019fnb}, \textit{black universes}~\cite{Bronnikov:2006fu}, or \textit{black bounces}~\cite{Simpson:2018tsi}. In these geometries, the presence of a throat corresponds to a defocusing of geodesics and typically allows for a resolution of the singularity. We will show that this is indeed the case for our spacetime, which nevertheless exhibits specific properties as it is obtained as a solution of the EOM rather than from a prescribed metric ansatz.

\begin{figure}[t!]
	\centering
	\includegraphics[width=0.85\columnwidth]{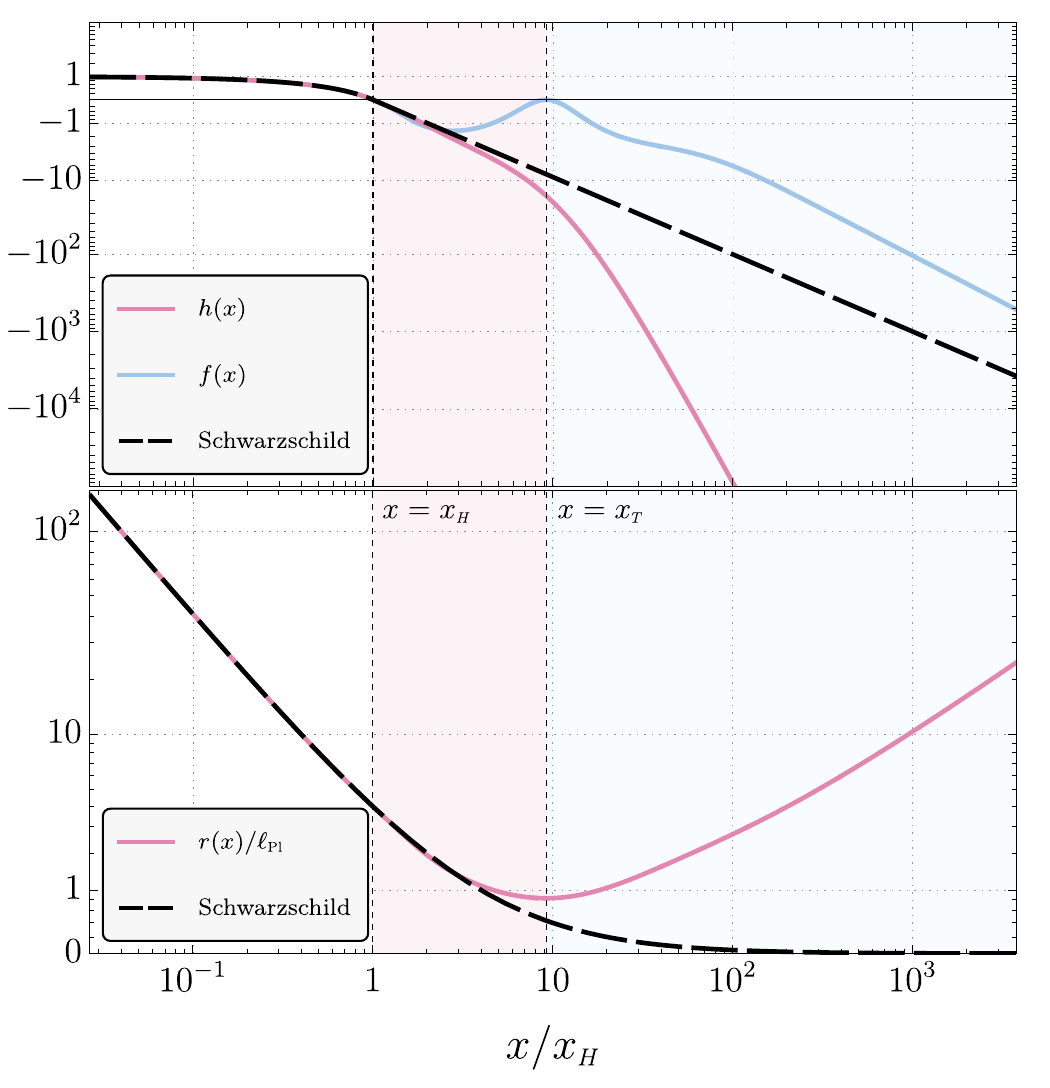}
	\caption{Metric of a one-way hidden wormhole with mass $M=2m_{\rm Pl}$ and Goroff-Sagnotti coupling $\lambda=1$. The top panel shows the metric functions $h$ and $f$ of the Schwarzschild ansatz as functions of the conformal-to-Kundt variable~$x$, while the bottom panel shows the radius $r$ as a function of~$x$. The region between the event horizon and the throat, both marked by vertical dashed lines, is highlighted in pink, while the region beyond the throat in the cosmological patch is highlighted in light blue. The dashed black curve indicates the Schwarzschild solution.}
\label{fig:metric}
\end{figure}

We show a representative example of this class of solutions in Fig.~\ref{fig:metric}. The presence of a wormhole throat makes the metric functions multivalued when expressed in terms of the 2-sphere radius $r$. We thus plot the Schwarzschild metric functions $h$ and $f$ as functions of the conformal-to-Kundt coordinate $x$, together with the radius as a function of $x$. The metric closely resembles the Schwarzschild solution throughout the exterior region, but departs significantly from it inside the event horizon. In particular, the radius reaches a minimum at the throat, beyond which both metric functions go to infinity.

To have a better understanding of the interior, it is useful to introduce the coordinate
\begin{equation}
    {\rm d}\tau= \frac{{\rm d}r}{\sqrt{-f(r)}}  \,.
\end{equation}
The metric in Eq.~\eqref{eq:static_patch} acquires a cosmological-like form:
\begin{equation}\label{eq:cosmological_patch}
    {\rm d}s^2=-{\rm d}\tau^2+ a_\parallel(\tau)^2 {\rm d}\rho^2  + a_\perp(\tau)^2 {\rm d}\Omega_2^2 \,,
\end{equation}
where we defined  $a_\parallel(\tau)^2=-h\left(r(\tau)\right)$, and we relabeled $t\to \rho$ and $r(\tau)\to a_\perp(\tau)$. We will refer to Eq.~\eqref{eq:cosmological_patch} as the \qu{cosmological patch}. One can explicitly check that this metric is regular at the wormhole throat, where $f(r)$ vanishes linearly. In fact, near $r = r_{\scriptstyle{T}}$, we have $f(r) \simeq - f_{\scriptstyle{T}} (r-r_{\scriptstyle{T}})$, with $f_{\scriptstyle{T}}>0$,  and hence $r(\tau) \simeq r_T+ f_{\scriptstyle{T}}\tau^2/4$, showing that the radius has a minimum at the throat.

This patch is particularly useful for understanding the limit $r\to \infty$ beyond the wormhole throat. The asymptotic leading behaviors of $f$ and $h$ are
\begin{equation} \label{eq:inside_asymptotics}
    f(r) \simeq -H_0^2 r^2 \,, \qquad h(r) \simeq - \frac{a_{\parallel,0}^2}{a_{\perp,0}^\alpha} r^\alpha \,,
\end{equation}
where $\alpha \simeq 7$, $H_0 \simeq 1.5 \lambda^{-1/4}$ are found using the Frobenius method, and are shown in the Supplemental material, while $a_{\parallel,0},a_{\perp,0}>0$ are free parameters. Therefore $a_\perp(\tau)=a_{\perp,0}\mathrm{e}^{H_0\tau}$, and the line element becomes
\begin{equation} 
    {\rm d}s^2 \simeq -{\rm d}\tau^2+ a_{\parallel,0}^2 \mathrm{e}^{\alpha H_0 \tau} {\rm d}\rho^2  + a_{\perp,0}^2 e^{2H_0\tau} {\rm d} \Omega_2^2 \,.
\end{equation}
This metric corresponds to a Kantowski-Sachs cosmology~\cite{Kompaneets:1965,Kantowski:1966te,Tsagas:2007yx}, in which $\tau \to \infty$ as $r \to \infty$. We note that similar interior structures have been found in different models of regular black holes~\cite{Modesto:2004wm,Peltola:2008pa,Modesto:2008im,Brandenberger:2021ken}.

The asymptotic scaling $f(r) \sim r^2$ ensures that the Kretschmann scalar $\mathcal K \equiv R_{\mu \nu \rho \sigma}R^{\mu \nu \rho \sigma}$ remains finite:
\begin{align} 
   \mathcal{K} \to \left(\frac{\alpha^4}{4}+2\alpha^2+12\right)H_0^4,\quad \text{as}\ r\to\infty\,.
\end{align}
Since the spacetime is spherically symmetric, this also ensures the finiteness of all algebraic curvature invariants~\cite{Bronnikov:2012wsj}.  Specifically, the Goroff-Sagnotti term in our action approaches $C^3 \to \left(\alpha(\alpha-2)H_0^2\right)^3/144$, so that Eq.~\eqref{eq:action}, evaluated on our solution, does not entail any divergence. More details on the behavior of curvature invariants are given in the Supplemental Material. This suggests that the spacetime might be free from singularities. We next investigate whether it is also geodesically complete.

\paragraph*{Geodesics completeness and causal structure.}

\begin{figure}[t!]
	\centering
	\includegraphics[width=0.85\columnwidth]{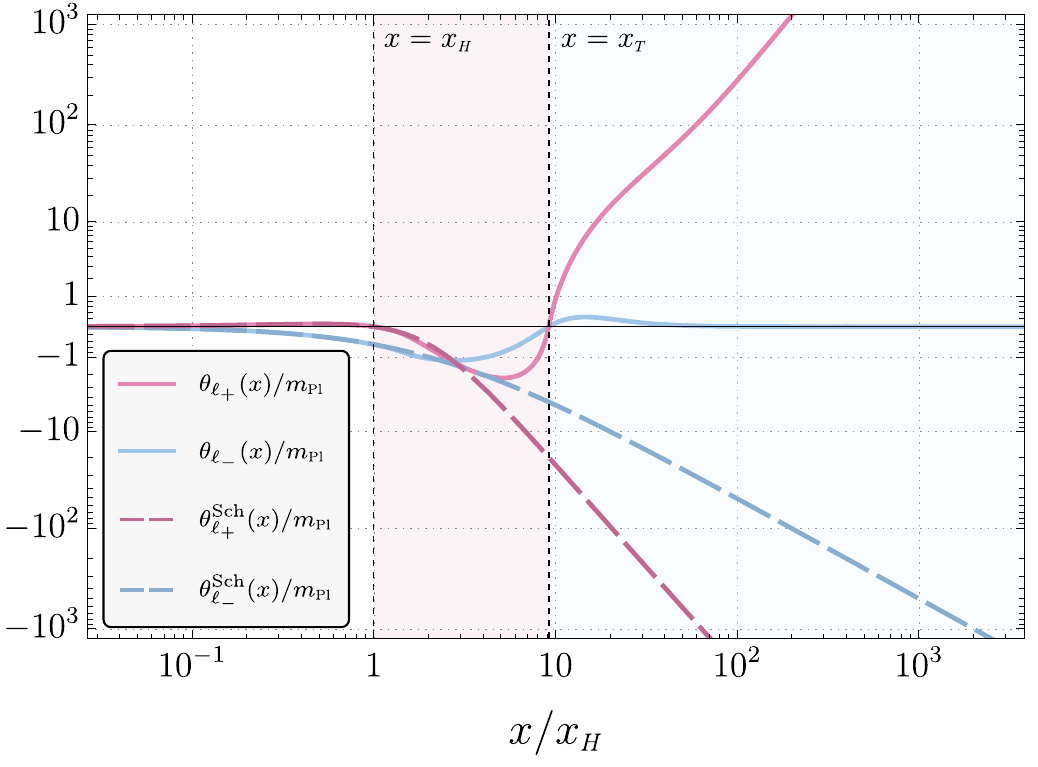}
	\caption{Expansions of the one-way hidden wormhole shown in Fig.~\ref{fig:metric} as functions of the conformal-to-Kundt variable $x$. The region between the event horizon and the throat, both marked by vertical dashed lines, is highlighted in pink, while the region beyond the throat in the cosmological patch is highlighted in light blue. In the exterior region, the expansion of outgoing geodesics is positive, whereas that of ingoing geodesics is negative. Both expansions are negative between the event horizon and the throat, and positive beyond the throat. The dashed curves indicate the expansions in the Schwarzschild case.}
\label{fig:expansions}
\end{figure}

Instead of working within the static patch~\eqref{eq:static_patch}, we adopt the globally-defined conformal-to-Kundt ansatz~\eqref{eq:conf_kundt}. The metric is regular everywhere away from $\Omega=0$ and well-suited to describe our solution. First, let us focus on the radial null geodesics. A vanishing line element in Eq.~\eqref{eq:conf_kundt} implies ${\rm d}v=0$ and $2{\rm d}x= \mathcal H(x){\rm d}v$. The two correspondent tangent null vectors are
\begin{equation} \label{eq:radial_geodesics}
    \ell_-^\mu \partial_\mu= \frac{1}{\Omega(x)^2} \partial_x \,, \qquad \ell_+^\mu \partial_\mu= \partial_v + \frac{\mathcal H(x)}{2} \partial_x\,.
\end{equation}
The overall factors are chosen in such a way that $\ell_+ \cdot \ell_-=-1$ and $\ell_-^\mu$ is affinely parametrized, $\ell_-^\mu \nabla_\mu \ell_-^\nu=0$. The subscript $+$ ($-$) indicates the outgoing (ingoing) radial geodesic. The corresponding expansions are defined as~\cite{Poisson:2009pwt}
\begin{equation} \label{eq:expansions}
    \theta_{\ell_\pm}= \frac{1}{\sqrt{h}} \mathcal L_{\ell_\pm} \sqrt{h}= \frac{2}{\Omega(x)} \ell_\pm^\mu \partial_\mu \Omega(x)\,,
\end{equation}
with $h=\det(h_{\mu \nu})$ being the determinant of the induced metric on the submanifold locally orthogonal to $\{ \ell_-^\mu\, ,\ell_+^\mu \}$, namely $h^{\mu \nu}=g^{\mu \nu}-2\ell_+^{(\mu}\ell_-^{\nu)}$. Explicitly, one has
\begin{equation} \label{eq:expansions_formula}
    \theta_{\ell_-}= \frac{2 \Omega'(x)}{\Omega(x)^3}\,, \qquad \theta_{\ell_+}= \frac{\mathcal H(x) \Omega'(x)}{\Omega(x)} \,.
\end{equation}
At the horizon, only the outgoing expansion changes sign, and $\theta_{\ell_+}(x_H)=0$ while $\theta_{\ell_-}(x_H)\ne0$. At the wormhole throat, instead, both expansions vanish $\theta_{\ell_\pm}(x_T)=0$, and swap sign on the two sides. The behavior of the two expansions across the different regions is depicted in Fig.~\ref{fig:expansions}.

The wormhole throat represents a local minimum for both the expansions. In fact
\begin{equation} \label{eq:expansions_troath}
    \ell_-^\mu \partial_\mu\theta_{\ell_-} \Biggr|_{x_T}= \frac{2 \Omega''}{\Omega^5}\Biggr|_{x_T}\,, \quad \ell_+^\mu \partial_\mu\theta_{\ell_+}\Biggr|_{x_T}= \frac{\mathcal H \Omega''}{\Omega}\Biggr|_{x_T} \,,
\end{equation}
and, at the throat $\mathcal H>0$, $\Omega>0$, and $\Omega''>0$, making both expressions in Eq.~\eqref{eq:expansions_troath} positive-valued. Note that the convexity of $\Omega$ characterizes the wormhole throat as a hypersurface with sections of minimal radius $r=\Omega(x_T)$ (see Fig.~\ref{fig:metric}), or, equivalently, of minimal area $\mathcal A(x_T)=4 \pi \Omega(x_T)^2$. 

Incidentally, this shows violations of the null convergence condition (NCC). The Raychaudhuri equation for spherical null congruences is~\cite{Poisson:2009pwt}
\begin{equation} \label{eq:Raychaudhuri_ell_-}
   \ell_-^\mu \partial_\mu\theta_{\ell_-}= \frac{{\rm d}\theta_{\ell_-}}{{\rm d}\sigma}=- \frac{\theta_{\ell_-}^2}{2} - R_{\mu \nu} \ell_-^\mu \ell_-^\mu\,,
\end{equation}
where $\sigma$ is an affine parameter for the ingoing geodesic. At the throat, this gives
\begin{equation} \label{eq:Raychaudhuri_ell_-_2}
   0<\ell_-^\mu \partial_\mu\theta_{\ell_-} \Biggr|_{x_T}= \frac{{\rm d}\theta_{\ell_-}}{{\rm d}\sigma}\Biggr|_{x_T}= - R_{\mu \nu} \ell_-^\mu \ell_-^\mu\Biggr|_{x_T}\,,
\end{equation}
explicitly violating the NCC. The same conclusion holds for $\theta_{\ell_+}$. Such a violation evades the assumption of Penrose's singularity theorem~\cite{Penrose:1964wq}: in our solution, both the ingoing and the outgoing congruences focus until the minimal radius $\Omega(x_T)$, before entering a defocusing region\footnote{A possible interpretation of the NCC violation is the following. In vacuum GR the NCC is saturated, $R_{\mu\nu}k^\mu k^\nu=0$, so the defocusing at the throat requires a Ricci curvature of the same order as the total curvature there: the solution is not a small deformation of a vacuum solution of GR. This signals a breakdown of the EFT regime around $r_T$. In this regime, the effective stress-energy of the Goroff-Sagnotti term can violate the NCC (see also~\cite{Belenchia:2016bvb}).}.

To establish the geodesic completeness of the spacetime, we turn to the cosmological patch~\eqref{eq:cosmological_patch}. We consider a radial geodesic $k^\mu$, normalized such that $k^\mu k_\mu= \epsilon$, with $\epsilon=0$ ($\epsilon=-1$) for null (timelike) curves. We parametrize the geodesic with an affine parameter~$\sigma$. The Killing vector $\chi^\mu \partial_\mu= \partial_\rho$ provides a notion of conserved quantity along the geodesic~\cite{Poisson:2009pwt}, namely $\chi \cdot k= E=H^2 {\rm d}\rho/{\rm d} \sigma$ and
\begin{equation} 
   \left( \frac{{\rm d}\tau}{{\rm d}\sigma} \right)^2= \frac{E}{a_{\parallel}(\tau)^2}- \epsilon \simeq \frac{E}{a_{\parallel,0}^2}  e^{-\alpha H_0 \tau} - \epsilon \,.
\end{equation}
For $\tau \to \infty$ the affine parameter behaves as
\begin{align} 
   &\sigma (\tau) \simeq \tau \,, \qquad &\epsilon=-1 \,, \\
   &\sigma(\tau) \simeq \frac{a_{\parallel,0}^2}{E} e^{\alpha H_0 \tau/2 } \,, \qquad &\epsilon=0\,.
\end{align}
In both cases $\sigma \to \infty$ as $\tau \to \infty$. That is, the geodesics employ an infinite amount of affine parameter (or proper time, if timelike) to reach the other asymptotic region, which is therefore geodesically complete.

\begin{figure}[t!]
	\centering
	\includegraphics[width=0.85\columnwidth]{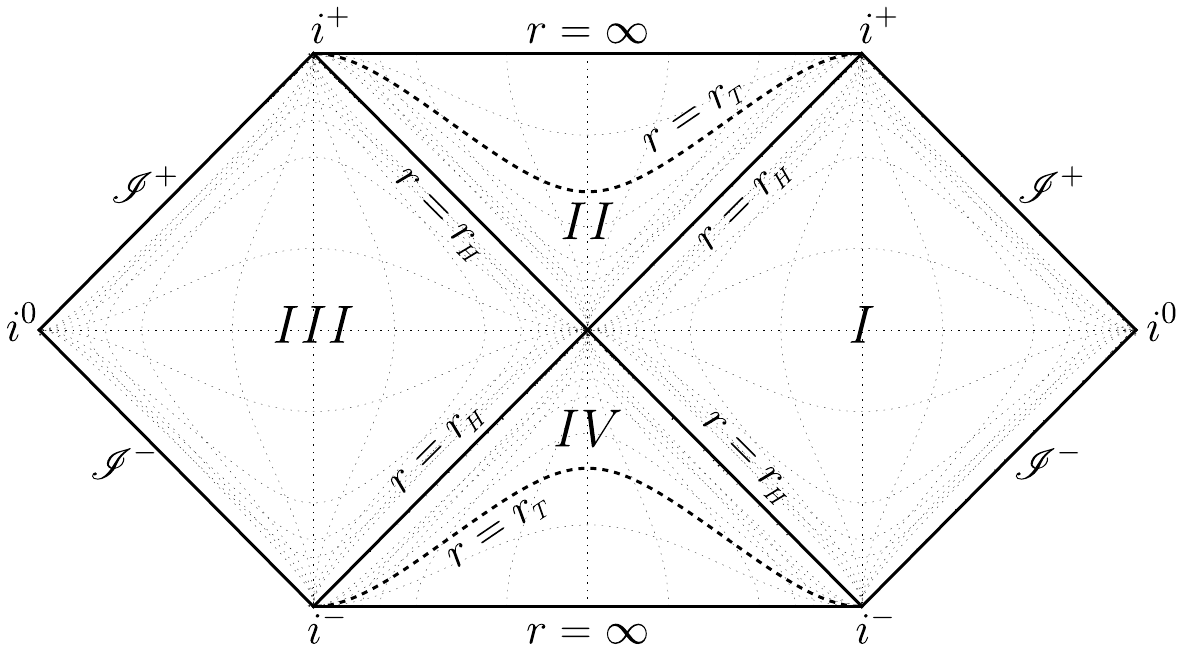}
	\caption{Penrose diagram of a maximally extended one-way hidden wormhole: beyond the event horizon, a spacelike wormhole throat leads to a future spacelike infinity, as in de Sitter spacetime, replacing the Schwarzschild singularity.}
\label{fig:penrose}
\end{figure}

Introducing the conformal time
\begin{equation} \label{eq:conformal_time}
    {\rm d} \eta= -\frac{{\rm d} \tau}{a_{\parallel,0}}e^{-\alpha H_0\tau/2} \,,
\end{equation}
one can cast the metric into the form
\begin{equation} \label{eq:conformal_time_metric}
    {\rm d} s^2= a^2_{\parallel,0}c(\eta)^2 \left(- {\rm d} \eta^2 + {\rm d}\rho^2 +   \frac{a_{\perp,0}^2}{ a_{\parallel,0}^2}c(\eta)^{\frac{4-2\alpha}{\alpha}} {\rm d}\Omega^2_2 \right) \,,
\end{equation}
with $c(\eta)=2/(\alpha H_0 \eta a_{\parallel,0})$.  Integrating Eq.~\eqref{eq:conformal_time}, one can verify that the surface $\tau \to \infty$ corresponds to a finite limit $\lim_{\tau \to \infty}\eta(\tau) $, that is, a surface of constant conformal time. Therefore, the inner asymptotic region has a spacelike character\footnote{Note that, strictly speaking, a smooth compactification of the four dimensional spacetime does not exist in the sense of~\cite{Penrose:1962ij}. In fact, let us pick a conformal factor $\omega(x)$, such that $\omega=0$ defines the boundary at $\tau \to \infty$. The rescaled metric would be $g_{\mu \nu}= \omega^2 \bar{g}_{\mu \nu}$. If $\bar g_{\mu \nu}$ was smooth, then the associated Weyl tensor $\bar C^2$ would be regular and one would have $C^2= \omega^4 \bar C^2 \to 0$ as $\tau \to \infty$. However, $C^2 \to (\alpha-2) \ne 0$ for our Kantowski-Sachs cosmology. Therefore, the diagram in Fig.~\ref{fig:penrose} corresponds to the radial compactification of the spacetime, while the two-sections in the $\{\theta, \phi \}$-subspace become degenerate.}. The Penrose diagram of such a spacetime is depicted in Fig.~\ref{fig:penrose}. The causal structure resembles the one-way hidden wormhole of~\cite{Carballo-Rubio:2019fnb}, but with a non-flat asymptotic region inside the horizon. The asymptotically flat region is connected, through an event horizon and a wormhole throat, to a spatially homogeneous, anisotropic spacetime of the Kantowski-Sachs type with a spacelike boundary at $r \to \infty$. 

All in all, our solution provides a geodesically complete version of classical black holes, in which a wormhole connects two inequivalent asymptotic regions. Remarkably, there are no inner horizons and the throat is screened by an event horizon, thus avoiding the typical instabilities of regular alternatives to classical black holes.

\paragraph*{Conclusions.}

This Letter derived one-way hidden wormholes as solutions of the EOM of a four-dimensional gravitational action motivated by EFT. The geometry is characterized by an outside event horizon and an interior wormhole throat. Crossing the throat leads not to a second asymptotically flat region, but to a cosmological spacetime of the Kantowski-Sachs type. The resulting geometry is geodesically complete and free of curvature singularities, and possesses no inner horizon. The emergence of a regular interior from the modified field equations provides an explicit realization of how EFT-motivated higher-curvature corrections can profoundly modify black hole interiors, including the global causal structure of the spacetime, while leaving the asymptotic geometry essentially unchanged. Remarkably, the solution avoids the standard mass-inflation mechanism associated with the inner horizons of conventional regular black holes, as well as other types of known instabilities~\cite{Carballo-Rubio:2025fnc}. While no instabilities are known for one-way hidden wormholes, this does not prove the full dynamical stability of the solution, which will need to be investigated in future works.

An intriguing question concerns the global causal structure of our solution and, in particular, its possible dynamical formation. The one-way hidden wormhole geometry features a nontrivial global structure whose formation from conventional gravitational collapse may require a change in spacetime topology. This raises a question that goes beyond the existence of regular solutions to the equations of an effective theory: can such geometries actually form within a quantum field theory of gravity in four dimensions? A conventional EFT description of the metric, formulated on a fixed spacetime manifold, does not by itself provide a framework for topology-changing processes. Moreover, classical results constrain topology change under suitable assumptions on causality and spacetime regularity. A consistent description of the formation of these geometries may therefore require an underlying theory of quantum gravity in which changes of spacetime topology can be accommodated~\cite{Hawking:1978pog,Gibbons:1991tp,Marolf:2020xie,Balasubramanian:2022lnw,Gesteau:2024gzf,Basile:2025zjc,McNamara:2026isz}. In this sense, our results provide a concrete setting in which to investigate the interplay between higher-curvature gravitational dynamics, the resolution of black hole singularities, and the global structure of quantum spacetime.

More broadly, our findings highlight that the leading higher-curvature corrections to GR can yield black holes with qualitatively new causal and geometric properties. Understanding their dynamical formation and stability, as well as their implications for black hole phenomenology, provides an exciting direction for exploring the low-energy manifestations of quantum gravity.

\begin{acknowledgments}
The authors thank I. Basile, F. Di Filippo, S. Liberati, and V. Vellucci for discussions, and I. Basile and B. Knorr for comments on the manuscript. The research of F.D.P. and A.P. is supported by a research grant (VIL60819) from VILLUM FONDEN. The Center of Gravity is a Center of Excellence funded by the Danish National Research Foundation under grant No. 184. This project has received funding from the European Union's Horizon 2020 research and innovation programme under the Marie Sklodowska-Curie grant agreement No 101299389 and No 101131233. This work is supported by Simons Foundation International~\cite{sfi} and the Simons Foundation~\cite{sf} through Simons Foundation grant SFI-MPS-BH-00012593-11. F.D.P. acknowledges Perimeter Institute for Theoretical Physics for their hospitality during the early stages of this work. Research at Perimeter Institute is supported in part by the Government of Canada through the Department of Innovation, Science and Economic Development Canada and by the Province of Ontario through the Ministry of Colleges and Universities. 
\end{acknowledgments}

\vfill

\bibliographystyle{apsrev4-1}
\bibliography{references.bib}

\newpage

\onecolumngrid
\section*{Supplemental Material}
\twocolumngrid

In this appendix we provide details on the EOM, the asymptotic and near-horizon expansions, and the numerical implementation to find the one-way hidden wormhole.

The EOM in general tensor form are
\begin{equation}\label{eq|eom_tensor}
\begin{split}
    &\,\frac{1}{16\pi G}\bigg(R_{\mu\nu}-\frac{1}{2}Rg_{\mu\nu}\bigg)+\frac{\lambda}{8}\bigg(3{C_{\mu}}^{\alpha\beta\gamma}C_{\nu\alpha\delta\lambda}{C^{\delta\lambda}}_{\beta\gamma}\\
    &\,-\frac{1}{2}{C_{\alpha\beta}}^{\gamma\delta}{C_{\gamma\delta}}^{\lambda\eta}{C_{\lambda\eta}}^{\alpha\beta}g_{\mu\nu}+R_{\mu\nu}C^{\alpha\beta\gamma\delta}C_{\alpha\beta\gamma\delta}\\
    &\,-R{C_\mu}^{\alpha\beta\gamma}C_{\nu\alpha\beta\gamma}+3R^{\alpha\beta}{C_{\mu\alpha}}^{\gamma\delta}C_{\nu\beta\gamma\delta}\\
    &\,-3R_{\alpha (\mu}C_{\nu)\beta\gamma\delta}C^{\alpha\beta\gamma\delta}+6\nabla_{\alpha}\nabla_{(\mu}\Big(C_{\nu)\beta\gamma\delta}C^{\alpha\beta\gamma\delta}\Big)\\
    &\,+6\nabla_\alpha\nabla_\beta\Big(C_{\mu\alpha\gamma\delta}{C_{\nu\beta}}^{\gamma\delta}\Big)-3g_{\mu\nu}\nabla_\alpha\nabla_\beta\Big(C^{\alpha\gamma\delta\lambda}{C^{\beta}}_{\gamma\delta\lambda}\Big)\\
    &\,-3\nabla^\alpha\nabla_\alpha\Big(C_{\mu\beta\gamma\delta}{C_{\nu}}^{\beta\gamma\delta}\Big)-\nabla_{\mu}\nabla_{\nu}\Big(C^{\alpha\beta\gamma\delta}C_{\alpha\beta\gamma\delta}\Big)\\
    &\,+g_{\mu\nu}\nabla^\alpha\nabla_\alpha\Big(C^{\beta\gamma\delta\lambda}C_{\beta\gamma\delta\lambda}\Big)\bigg)=0\,,
\end{split}
\end{equation}
where round brackets in the indices indicate symmetrization. In conformal-to-Kundt coordinates they can be reduced to an autonomous system of two differential equations~\cite{Daas:2023axu}
\begin{equation} \label{eq:eom}
\begin{split}
    \mathcal{H}'''=&\ \frac{\left(2+\mathcal{H}''\right)\left(3\mathcal{H}'\Omega'+\left(\mathcal{H}''-1\right)\Omega\right)}{3\mathcal{H}'\Omega}\\
    &\ -\frac{48\Omega^2\left(\mathcal{H}'\Omega'\Omega+3\mathcal{H}\Omega'^2+\Omega^2\right)}{\lambda\mathcal{H}'\left(2+\mathcal{H}''\right)}\,,\\
    \Omega''=&\ \frac{\left(2+\mathcal{H}''\right)\left(\lambda\left(2+\mathcal{H}''\right)^2-144\Omega^4\right)}{864 \mathcal{H}\Omega^3}-\frac{\mathcal{H}'\Omega'}{\mathcal{H}}\,.
\end{split}
\end{equation}

Since an event horizon is located at $\mathcal{H}=0$, which is a singular point of the EOM, we investigate black hole solutions by implementing a shooting method between an asymptotically flat region and an event horizon. To have greater control over the form of the metric in the asymptotically flat region, we start from a power series in $r^{-1}$ in Schwarzschild coordinates:
\begin{equation} \label{eq:asym_flat}
\begin{split}
    h(r)\sim&\ 1-\frac{2 m}{r}+\frac{3\lambda m^2}{r^6}-\frac{4\lambda m^3}{r^7}+O\left(\frac{\lambda^2 m^3}{r^{11}}\right)\,,\\
    f(r)\sim&\ 1-\frac{2 m}{r}+\frac{9\lambda m^2}{r^6}-\frac{16\lambda m^3}{r^7}+O\left(\frac{\lambda^2 m^3}{r^{11}}\right)\,,
\end{split}
\end{equation}
where $m=GM$, with $M$ the ADM mass. More precisely, we have checked that, up to the $r^{-32}$ term, the series follows the structure
\begin{equation}\label{eq:asyntot_series}
h,f\sim 1+\displaystyle\sum_{i=0}^{n}\displaystyle\sum_{j=i}^{2i}c_{i,j}\left(\frac{\lambda}{r^4}\right)^i\left(\frac{m}{r}\right)^{j+1}\,,
\end{equation}
where the $c_{i,j}$ are numerical coefficients that differ for $h$ and $f$. Using this expansion, we calculate the value of the coordinate $x_0$ and the values of $\mathcal{H},\mathcal{H}',\mathcal{H}'',\Omega$, and $\Omega'$ at a starting radius $r_0\gg 2m$, which we choose to be $r_0=150$ in Planck units. At the horizon, we directly expand the functions $\mathcal{H}$ and $\Omega$ in power series:
\begin{equation}\label{eq:hor_series}
\begin{split}
    \mathcal{H}(x)\sim&\ \mathcal{H}_1(x-x_{\scriptstyle{H}})+\mathcal{H}_2(x-x_{\scriptstyle{H}})^2+\displaystyle\sum_{i=3}^{n}\mathcal{H}_i(x-x_{\scriptstyle{H}})^i\,,\\
    \Omega(x)\sim&\ r_{\scriptstyle{H}}+\displaystyle\sum_{i=1}^{n}\Omega_i(x-x_{\scriptstyle{H}})^i\,,
\end{split}\raisetag{0.4cm}
\end{equation}
where the coefficients $\mathcal{H}_i$ and $\Omega_i$ are functions of $r_{\scriptstyle{H}},\mathcal{H}_1$, and $\mathcal{H}_2$. In our calculations, we use the expansions up to fourth order for both functions and their derivatives, and integrate from a point satisfying $x-x_{\scriptstyle{H}}=10^{-9}$.

We integrate from both the asymptotic region and the horizon up to a fitting point $x_{\scriptstyle{H}}+10^{-4}$, where we match the two solutions using a root-finding algorithm. This allows us to determine the appropriate values of $x_{\scriptstyle{H}},r_{\scriptstyle{H}},\mathcal{H}_1$, and $\mathcal{H}_2$ for given values of the asymptotic mass $m$ and the Goroff-Sagnotti coupling $\lambda$. The numerical implementation is performed using Wolfram \texttt{Mathematica}. Both the numerical integration and the root-finding algorithm use a \texttt{PrecisionGoal} and an \texttt{AccuracyGoal} of 16 digits, with a \texttt{WorkingPrecision} of 32 digits. We have verified that the numerical results are stable under these settings.

Having determined the parameters at the horizon, we can start a new integration in the interior region, for which we use a \texttt{WorkingPrecision} of 48 digits. This integration does not require additional machinery, as the only other possible singular point is at the center of the geometry, where $\Omega=0$. The other potentially singular points, corresponding to $\mathcal{H}'=0$ and $2+\mathcal{H}''=0$, do not affect the numerical integration along the solutions we consider. Nevertheless, the functions can exhibit very stiff behavior, and the integration does not always converge, particularly for large masses. 

In particular, from a preliminary analysis of the range $\lambda\in[10^{-3},10^3]$, we found solutions with an event horizon for masses $M\gtrsim\lambda^{1/4}m_{\text{Pl}}$, while numerical stability, and thus the existence of one-way hidden wormholes, was found consistently only for masses $M\lesssim2.5\lambda^{1/4}m_{\text{Pl}}$. However, a systematic investigation of the parameter space is postponed to future work.

We investigated the leading-order behavior of the solutions at spatial infinity using a Frobenius expansion~\cite{Stelle:1977ry,Pravda:2020zno,Giacchini:2024exc}, starting from the power series ansatz
\begin{equation}
    h(r)\sim r^\alpha\displaystyle\sum_{i=0}^{n}h_i r^i, \qquad f(r)\sim r^\beta\displaystyle\sum_{i=0}^{n}f_i r^i\,.
\end{equation}
Finiteness of the curvature invariants requires $\beta \leq 2$. For $\beta <2$, all curvature invariants vanish asymptotically, and the only solution is asymptotically flat. It is therefore described by the expansion in eq.~\eqref{eq:asym_flat}. The case $\beta=2$, for which the curvature invariants approach finite, nonzero values, reduces the field equations to a set of algebraic equations at cubic order. The only real solutions have
\begin{equation}
\begin{split}
        \alpha=&\, \frac{1}{3} \left(4+\sqrt[3]{1288-54 \sqrt{502}}+\frac{29\ 2^{2/3}}{\sqrt[3]{644-27 \sqrt{502}}}\right)\,,\\
        f_0=&\, \pm \frac{\lambda^{-1/2}}{9} \Bigg(\frac{841 \sqrt[3]{2}}{\left(644-27 \sqrt{502}\right)^{2/3}}+\frac{116\
   2^{2/3}}{\sqrt[3]{644-27 \sqrt{502}}}\\
   &+\,\frac{1}{2} \left(1288-54 \sqrt{502}\right)^{2/3}+291\\
   &+4 \sqrt[3]{1288-54 \sqrt{502}}\Bigg)^{1/2}\,,
\end{split}\raisetag{0.4cm}
\end{equation}
while $h_0$ remains a free parameter. Using the cosmological coordinates in eq.~\eqref{eq:cosmological_patch}, we identify $H_0=\sqrt{|f_0|}$. There are no solutions with $\beta>2$, and thus with divergent curvature invariants.

\begin{figure}[t!]
	\centering
	\includegraphics[width=0.85\columnwidth]{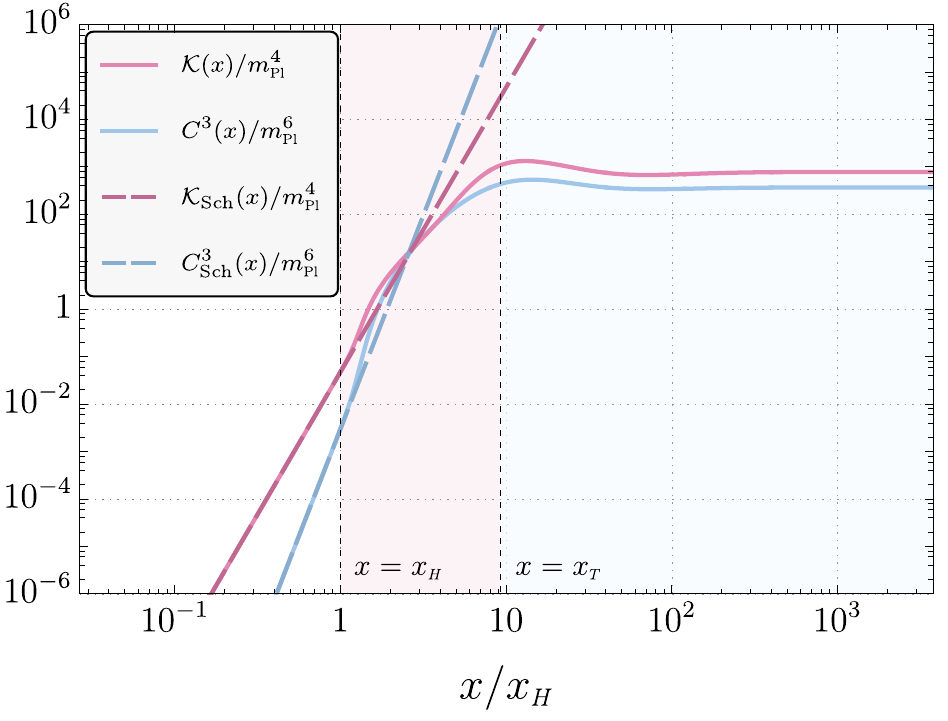}
	\caption{Kretschmann and Goroff-Sagnotti curvature invariants of the one-way hidden wormhole shown in Fig.~\ref{fig:metric} as functions of the conformal-to-Kundt variable $x$. The region between the event horizon and the throat, both marked by vertical dashed lines, is highlighted in pink, while the region beyond the throat in the cosmological patch is highlighted in light blue. The dashed curves indicate the expansions in the Schwarzschild case.}
\label{fig:invariants}
\end{figure}

Finally, we include for completeness a plot of the Kretschmann and Goroff-Sagnotti curvature invariants in Fig.~\ref{fig:invariants}, for $\lambda=1$. Both start from very small values, reach Planckian values after crossing the event horizon, attain a maximum just beyond the throat, and then approach constant values. By contrast, in the Schwarzschild metric, both invariants diverge as the origin is approached. The Ricci scalar is not shown in the plot, as it vanishes identically for the Schwarzschild solution and is proportional to the Goroff-Sagnotti invariant for the hidden wormhole. Indeed, the trace of the EOM reads
\begin{equation}
    -\frac{1}{16\pi G}R+\frac{\lambda}{8}C^3=0\,,
\end{equation}
and the two invariants are not independent for the solutions of the theory

\let\clearpage\relax 



\vfill

\end{document}